\documentclass[showpacs,prl,floatfix,twocolumn,amsmath,a4]{revtex4}
\usepackage{graphicx}

\begin{document} 

\title{Electronic correlations decimate the ferroelectric polarization of multiferroic HoMn$_2$O$_5$}

\author{Gianluca Giovannetti$^{1,2}$ and Jeroen van den Brink$^{1,3}$}

\address{
$^1$Institute Lorentz for Theoretical Physics, Leiden University, 
          P.O. Box 9506, 2300 RA Leiden, The Netherlands\\ 
$^2$Faculty of Science and Technology and MESA+ Research Institute, University of Twente,
            P.O. Box 217, 7500 AE Enschede, The Netherlands\\ 
$^3$Institute for Molecules and Materials, Radboud Universiteit Nijmegen,
P.O. Box 9010, 6500 GL Nijmegen, The Netherlands}

\begin{abstract} 
We show that  electronic correlations decimate the intrinsic ferroelectric polarization of the recently discovered class of multiferroic manganites RMn$_2$O$_5$, where R is a rare earth element. Such is manifest from {\it ab initio} bandstructure computations that account for the strong local Coulomb interactions between the manganese $3d$ electrons --the root of magnetism in these materials. When including these the computed electronic, magnetic and lattice structure of multiferroic HoMn$_2$O$_5$ results in an amplitude and direction of polarization that is in accordance with experiment. The microscopic mechanism behind the decimation is a near cancellation of the ionic polarization induced by ferroelectric lattice displacements and the electronic one caused by valence charge redistributions.  
\end{abstract}

\date{\today} 


\maketitle

{\it Introduction} Multiferroics, single phase compounds in which magnetism and ferroelectrity coexist, are rare~\cite{Hill,Khomskii,Mostovoy}. These materials, such as for instance the manganites RMn$_2$O$_5$ (R=Ho, Tb, Y, Eu, etc.)~\cite{Hur,Kagomiya,Chapon04,Alfonso,Chapon06,Hur04,Higashiyama05}, are currently of great interest because of the possibility to control {\it magnetic} properties by {\it electric} fields and {\it vice versa}. It was recently discovered that in for instance TbMn$_2$O$_5$ the magneto-electric coupling is so strong, that the electric polarization can be reversed by an external magnetic field~\cite{Hur}. This breakthrough can open new routes for the design of magneto-electric devices. 

From a fundamental point of view, however, these multiferroic manganites contain a puzzle. In regular, non-magnetic ferroelectrics the size of the macroscopic polarization $P$ computed by modern {\it ab initio} bandstructure methods agrees exceptionally well with the ones observed experimentally~\cite{Rabe07}. In the multiferroic manganites, however, state of the art {\it ab initio} computations predict a $P$ of around 1200 $nC/cm^2$ (Tb~\cite{Wang07,Wang_tbp} and Ho), whereas the experimentally observed values are more than an order of magnitude smaller ($P$=45, 65, 100, and 115 $nC/cm^2$ for Tb, Ho, Y and Eu, respectively~\cite{Hur,Kagomiya,Chapon04,Alfonso,Chapon06,Hur04,Higashiyama05}). The question arises whether this large discrepancy is due experimental artifacts, for instance the formation of ferroelectric domains, or due to an incompleteness in our understanding of the physical properties of these magnetic ferroelectric materials. The outcome of this puzzle is not only of fundamental interest, as large theoretical values of $P$ promise experimentalists a boost of polarization upon enhanced material quality, increasing the multiferroics' application potential.

We will show in this Letter that the small polarization is intrinsic and caused by electronic correlations. It arises because the two contributions to $P$, the ionic part from the lattice displacements and the electronic part from the valence electrons are opposite and almost canceling each other. In this way the electron-electron interactions drive a decimation of the resulting net polarization. We compute $P$ to be in close agreement with the experimental value only when the strong local Coulomb interactions between the manganese $3d$ electrons are accounted for. 

\begin{figure}
\centerline{\includegraphics[width=.8\columnwidth,angle=0]{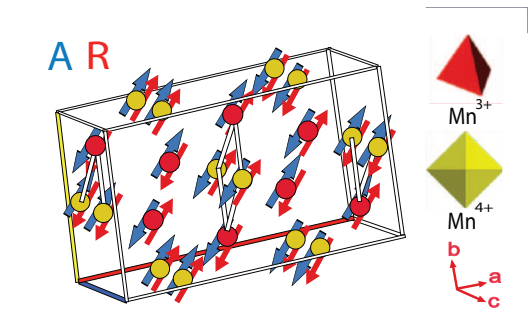}}
\caption{(Color online) Schematic view of the crystal structure of HoMn$_2$O$_5$ consting of connected Mn$^{4+}$O$_6$ octahedra (yellow) and Mn$^{3+}$O$_5$ pyramids (red). The magnetic structure of the ground state, labelled A, and its enantiomorphic counterpart, labelled R, are shown. The white bars connect Mn ions of the Mn$^{3+}$-Mn$^{4+}$-Mn$^{3+}$ structure along the $b$ axis.}
\label{fig1}
\end{figure}

{\it Structure of RMn$_2$O$_5$} In the following we will focus on the case R=Ho, but our conclusions are generic for this class of compounds. Neutron and X-ray diffraction studies show that these manganites have space group \emph{Pbam} but it is expected that in their multiferroic state the actual symmetry group is {\emph{Pb}2$_1$\emph{m}}, which allows for a macroscopic electric polarization along the $b$ axis~\cite{Alfonso,Kagomiya,Chapon04}. The orthorhombic $Pbam$ crystal structure of HoMn$_2$O$_5$ consists of connected Mn$^{4+}$O$_6$ octahedra and Mn$^{3+}$O$_5$ pyramids (see Fig.~\ref{fig1}). The octahedra share edges and form ribbons parallel to the $c$ axis. Adjacent ribbons are linked by pairs of corner-sharing pyramids. Below 38 K HoMn$_2$O$_5$  a commensurate magnetic structure develops with propagation vector ${\bf k}=({1 \over 2},0,{1 \over 4})$ and simultaneously the system becomes ferroelectric~\cite{Blake}. 

We expand upon previous {\it ab initio} calculations by including the very strong local Coulomb interactions between the manganese $3d$ electrons --the Hubbard $U$. We use the projector augmented-wave method (PAW) and plane wave basis sets as implemented in VASP~\cite{VASP}. Exchange and correlation are treated using the generalized gradient spin-density approximation \cite{PW91} (SGGA) and the SGGA+U method~\cite{PAW+U,Liechtenstein95}. We performed the SGGA+U calculations for $U$=4 and 8 eV and a Hund's rule exchange of $J_H=0.88$ eV for the Mn $d$-electrons, in accordance with values obtained from constrained density functional calculations on perovskite mangnites~\cite{Satpathy}. 

Starting from the experimental centrosymmetric \emph{Pbam} crystal structure we relax unit cell parameters and ionic positions both in the SGGA and SGGA+U schemes, allowing for the lower symmetry {\emph{Pb}2$_1$\emph{m}} structure to develop. The atomic positions are relaxed in a 2x1x1 magnetic super cell (containing 64 ions) along the magnetic $k_x$ direction~\cite{remark_computational_details}. We find that the experimental magnetic structure of HoMn$_2$O$_5$ (labeled by A in Fig.~\ref{fig1}) with {\emph{Pb}2$_1$\emph{m}} symmetry is indeed the magnetic groundstate. The calculated structural parameters for $U$=0 and $U$=8 are shown in Table~\ref{table1}.  They are in good agreement with both the experimental data and first principles electronic structure computations on TbMn$_2$O$_5$ without correlations~\cite{Wang07,Wang_tbp,Alfonso}. The ionic displacements are small but significant, in agreement with the fact that experimentally the low symmetry structure cannot directly be determined~\cite{Blake}.  In Table~\ref{table2} we report the computed band gap $\Delta$ and energy gain due to the ferroelectric distortion $\delta E_{FE}$ (see Fig.~\ref{fig:values_polarization}). 

\begin{table}
\begin{center}
\begin{tabular}{l | ccc | ccc}
\hline \hline
& & U=0.0 & &  & U=8.0  \\
$a$, $b$, $c$  & 14.5188 & 8.5271 & 5.6681 & 14.6847 & 8.5480 & 5.7858 \\
\hline
Ho$^{3+}$& 0.0693  & 0.1725 & 0      & 0.0690 & 0.1700 & 0      \\
         & 0.3190  & 0.3278 & 0      & 0.3178 & 0.3300 & 0      \\
Mn$^{4+}$& 0 & 0.5003 & 0.2559  & 0.9996 & 0.4997 & 0.2536 \\
Mn$^{3+}$& 0.2032  & 0.3530 & 0.5    & 0.2063 & 0.3482 & 0.5    \\
         & 0.4534  & 0.1485 & 0.5    & 0.4566 & 0.1533 & 0.5    \\
O$_1$    & 0.0004  & 0.0004 & 0.2702 & 0.0005 & 0.0004 & 0.2682 \\
O$_2$    & 0.0823  & 0.4447 & 0      & 0.0810 & 0.4415 & 0      \\
         & 0.3324  & 0.0549 & 0      & 0.3317 & 0.0584 & 0      \\0
O$_3$    & 0.0768  & 0.4305 & 0.5    & 0.0727 & 0.4242 & 0.5    \\
         & 0.3275  & 0.0674 & 0.5    & 0.3238 & 0.0736 & 0.5    \\
O$_4$    & 0.1983  & 0.2075 & 0.2446 & 0.1955 & 0.2057 & 0.2382 \\
         & 0.4471  & 0.2921 & 0.7571 & 0.4447 & 0.2942 & 0.7625 \\
\hline \hline
\end{tabular}
\end{center}
\caption{Computed structural parameters of HoMn$_2$O$_5$ in {\emph{Pb}2$_1$\emph{m}} crystal structure using SGGA, SGGA+U. Distances are in \AA ; atoms occupying equivalent Wyckoff positions are shown only once.}
\label{table1}
\end{table}

\begin{figure}
\centerline{\includegraphics[width=\columnwidth,angle=0]{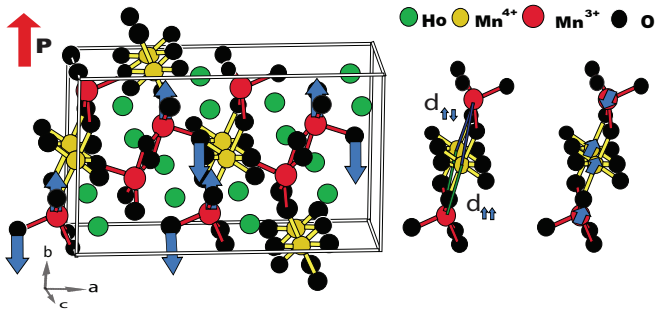}}
\caption{(Color online) Left: arrangement of the ions in the unit cell with arrows indicating the ionic displacements of Mn$^{3+}$ and O$_3$. Right: schematic view of the magnetic and charge ordered Mn$^{3+}_{\Uparrow}$-Mn$^{4+}_{\Uparrow}$-Mn$^{3+}_\Downarrow$ arrangement along the $b$ direction.}
\label{fig2}
\end{figure}

\begin{figure}                                                                                                                                              
\centerline{\includegraphics[width=.85\columnwidth,angle=0]{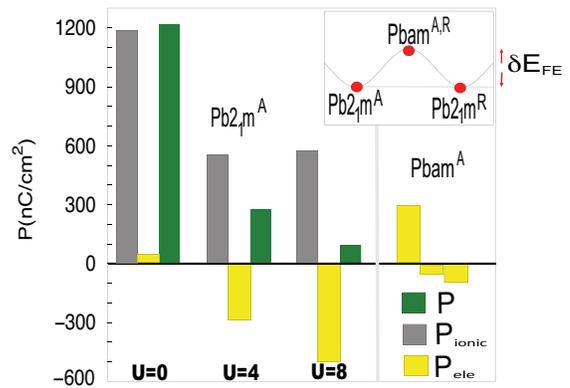}}
\caption{(Color online) Ionic, electronic and total polarization for different values of $U$ in the relaxed {\emph{Pb}2$_1$\emph{m}} structure (left) and $P_{ele}$ in the   \emph{Pbam} structure (right). Inset: ferroelectric energy gain, $\delta E_{FE}$.}
\label{fig:values_polarization}                                                                                                                                                
\end{figure}

{\it  Ferroelectric atomic displacements}
The the relaxation results in Mn$^{3+}$ and O$_3$ having significant atomic displacements along the $b$ direction, compared to which the displacements for other ions and in other directions are small. In Fig. \ref{fig2} the displacements of these two types of atoms are indicated. In the $a$ and $c$ direction the ionic displacements are mirror symmetric so that they will not contribute to developing a ferroelectric polarization. 

One qualitative difference between the relaxed unit cells obtained in SGGA and SGGA+U appears in the oxygen octahedra surrounding the Mn$^{4+}$ ions: SGGA calculations show that the Mn$^{4+}$ move along the $b$ direction and become off-centered (the bonds along the longest axis of the octahedron become 1.921 and 1.935 \AA, respectively) while switching on the Coulomb interaction $U$ results in a suppression of this off-centering. The effect is not unexpected as the inclusion of the electronic correlations increases the band gap significantly, rendering the electronic system more rigid and less susceptible to perturbations. The instability to off-centering is well known to be related to the hybridization of the transition metal $3d$ states with the oxygen $2p$ states~\cite{Hill,Filippetti}.  Increasing $U$ leads to a larger gap, a larger splitting between occupied oxygen $p$ and empty Mn $d$ states and therefore a smaller effective hybridization between the two.

Along the $b$-direction HoMn$_2$O$_5$ exhibits a charge and spin ordering that can schematically be denoted as a chain of Mn$^{3+}_{\Uparrow}$-Mn$^{4+}_{\Uparrow}$-Mn$^{3+}_\Downarrow$, see Fig.~\ref{fig2}.  In the undistorted \emph{Pbam} structure the distances d$_{\Uparrow\Uparrow}$ (between Mn$^{3+}_\Uparrow$ and Mn$^{4+}_\Uparrow$) and d$_{\Downarrow\Uparrow}$ (between Mn$^{3+}_\Downarrow$ and Mn$^{4+}_\Uparrow$) are the same. Relaxation reveals a shortening of distances between parallel spins Mn$^{3+}_{\Uparrow}$ and Mn$^{4+}_{\Uparrow}$ ions  --in the ferroelectric {\emph{Pb}2$_1$\emph{m}} structure $d_{\Uparrow\Uparrow} < d_{\Downarrow\Uparrow}$, which optimizes the double exchange energy~\cite{Mostovoy,Efremov04}. 

\begin{table}
\begin{ruledtabular}
\begin{tabular}{cccccc}
$U$  (eV) & $\Delta$ (eV) & $\delta E_{FE}$ (meV) &Mn$^{3+}_{\Uparrow}$ & Mn$^{4+}_{\Uparrow}$ & Mn$^{3+}_{\Downarrow}$ \\
0.0 & 0.5 & 26.4  &3.85  & 4.74  & 4.17  \\
4.0 & 1.6 & 12.1 &3.94  & 4.03  & 4.08  \\
8.0 & 1.6 & 18.6 &3.69  & 3.65  & 3.87 
\label{table2}
\end{tabular}
\end{ruledtabular}
\caption{Gap $\Delta$, energy gain of the ferroelectric state $\delta E_{FE}$ and Born effective charges of different manganese ions in multiferroic HoMn$_2$O$_5$ with {\emph{Pb}2$_1$\emph{m}} symmetry within SGGA and SGGA+U.}
\label{table2}
\end{table}

{\it Ionic and electronic polarization}
The total polarization $P$ for a given material is the sum of the ionic polarization $P_{ion}$ and electronic contribution one $P_{ele}$~\cite{Rabe07,Resta2}.  In our {\it ab initio} calculations the ionic contribution $P_{ion}$ is easily obtained by summing the product of the ionic displacements from the centrosymmetric to the ferrolectric structure with the nominal charge of the ions' rigid core. To calculate the electronic contribution $P_{ele}$ we use the Berry phase method developed by King-Smith and Vanderbilt within the PAW formalism \cite{BerryPhase1,BerryPhase2}.

First we consider the magnetically ordered {\it high symmetry} \emph{Pbam} phase of HoMn$_2$O$_5$, in which by definition $P_{ion}$=0. The electronic part to the polarization, however, is not required to vanish. In fact, the material is bound to have a magnetically induced ferroelectric polarization as it is a dislocated spin density wave system in which the center of symmetry of the magnetic and lattice structure do not coincide, providing a basic mechanism for multiferroicity~\cite{Betouras,Chapon06}. An equivalent point of view is that in this oxide symmetry allows for a purely electronic part to the magnetostriction~\cite{Mostovoy,Harris}. As a consequence the spin ordering induces a redistribution of charge on crystalographically inequivalent manganese sites. 

We find in the centrosymmetric A structure a resulting polarization of $P_{ele}=$284 nC/cm$^2$ along the $b$ axis for $U$=0. We checked that inverting all the spins,  producing the reversed (R) \emph{Pbam} structure, leads to the same polarization in opposite direction. A finite $U$ alters the ferroelectric charge redistribution and  gives rise to a polarization $P_{ele}$=-14/-81 nC/cm$^2$ for $U$=4/8 eV in the A spin structure~\cite{remark_invert}. The fact that $P_{ele}$ is induced by the magnetic superstructure is immediately clear from a computation on this system in the ferromagnetic state, in which we find all polarization to vanish.

In the results above it is remarkable that the electronic correlation effects induce a sign change of $P_{ele}$. This is a real effect caused by changes in electronic structure and is not related to geometric constraints in our calculations. A polarization flip is possible because symmetry considerations alone do not fix the sign of the magnetically induced polarization --the sign of the magneto-electric coupling. Thus an inversion of the polarization as a function of $U$ is symmetry allowed, just as a temperature induced sign change of the coupling is possible and indeed observed in some materials~\cite{Cheong_pc}. At the end of this paper we present the microscopic mechanism behind this correlation induced polarization flip.  

In the relaxed {\emph{Pb}2$_1$\emph{m}} structure also the ionic contribution to the polarization comes into play. We find $P_{ion}$=1193/546/576 nC/cm$^2$ for $U$=0/4/8, concomitant with an electronic polarization $P_{ele}$=12/-287/-493 nC/cm$^2$, resulting in a total polarization $P$=1205/259/82 nC/cm$^2$ in the magnetic A structure~\cite{remark_invert}. These values are shown in Fig.~\ref{fig:values_polarization}. The value of the polarization for $U$=8 is in very good agreement with experiment. From the computed values of $P_{ion}$ and $P_{ele}$ we conclude that the Hubbard $U$ causes a near cancellation of the electronic and ionic contributions to the polarization and effectively reduces the polarization in the multiferroic manganites by over an order of magnitude.

\begin{figure}
\centerline{\includegraphics[width=1\columnwidth,angle=0]{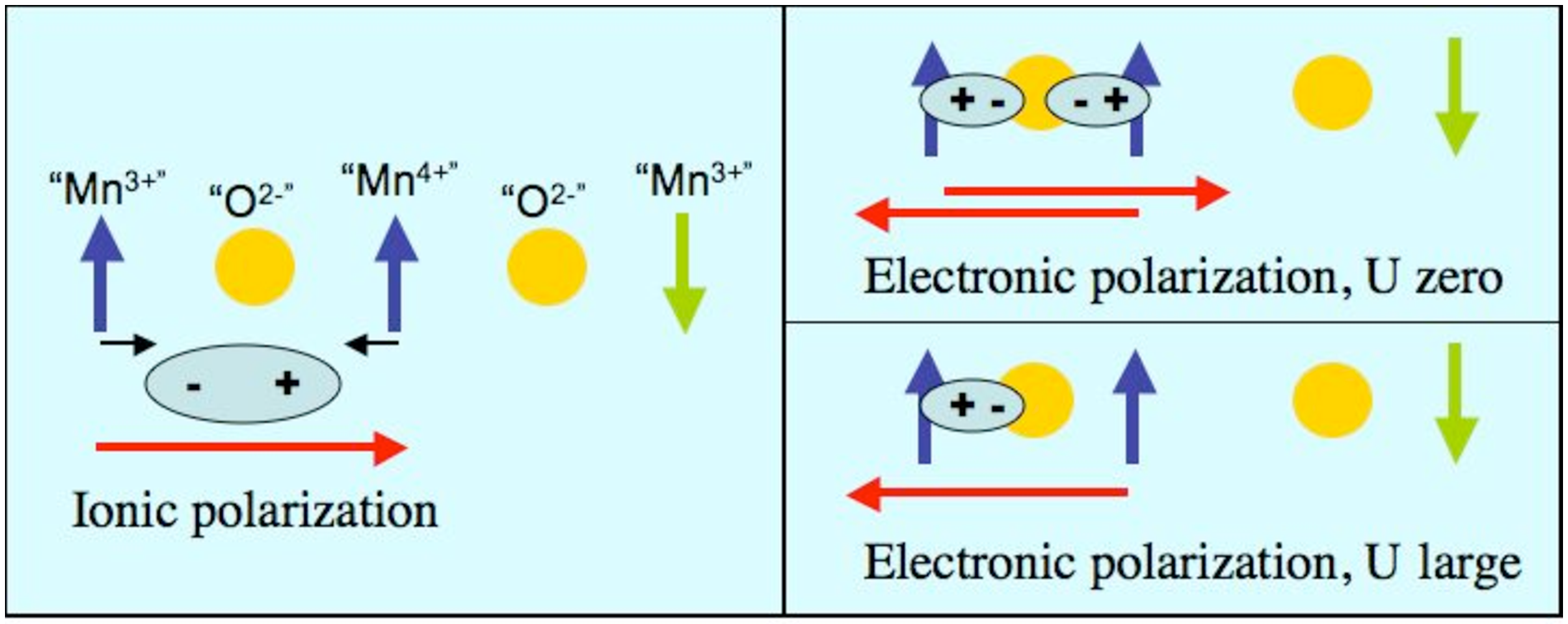}}
\vspace{-3cm}
\caption{(Color online) Schematic view of the two contributions to the ferroelectric polarization in HoMn$_2$O$_5$ in the uncorrelated ($U$=0) and strongly correlated limit (large $U$). In the latter the electronic polarization nearly cancels the ionic polarization. The labels "Mn$^{4+/3+}$" indicate Mn ions that have a valence of more/less than $3.5+$, respectively.}
\label{fig:schematic_polarization}
\end{figure}

{\it Origin of the near cancellation} 
The ionic contribution to the polarization is driven by fact that ($i$) the interatomic Mn distances depend on spin direction ($d_{\Uparrow\Uparrow} < d_{\Downarrow\Uparrow}$) and ($ii$) the valence of the two Mn ions approaching each other is different, see left panel of Fig.~\ref{fig:schematic_polarization}. Electronic correlations reduce $P_{ion}$ by a factor of two, due to the increased electronic rigidity which reduces atomic displacements, see Fig.~\ref{fig:values_polarization}. In spite of this correlation induced reduction the computed $P_{ion}$ is still about six times larger than the experimental polarization $P$.

The electronic polarization $P_{ele}$ arises from a reorganization of valence charges caused by both ferroelectric lattice distortions and magnetic ordering.  Both these cause changes in covalency, which in turn cause a flow of valence electron charge across the material. Therefore the effective charge that is displaced by a distortion can be much larger than just the bare ionic value. We computed this Born effective charges of the Mn ions along the Mn$^{3+}_\Uparrow$-Mn$^{4+}_\Uparrow$-Mn$^{3+}_\Downarrow$ direction in the ferroelectric {\emph{Pb}2$_1$\emph{m}} structure, see Table~\ref{table2}.

When $U$=0 all Born charges are larger than the nominal ones, indicative of the distortions inducing appreciable changes in covalency. The microscopic mechanism is that both the Mn$^{4+}_\Uparrow$ and Mn$^{3+}_\Uparrow$ ions transfer charge to the oxygen atoms connecting them when they move closer together, see Fig.~\ref{fig:schematic_polarization}. In this situation the electrons gain kinetic energy because they can hop between Mn$^{4+}_\Uparrow$ and Mn$^{3+}_\Uparrow$ without violating the on site Hund's rule~\cite{Efremov04}. The induced electronic polarizations (Fig.~\ref{fig:schematic_polarization}) are opposite and cancel each other, in agreement with the very small $P_{ele}$=12 nC/cm$^2$ that we find in {\emph{Pb}2$_1$\emph{m}} when $U$=0. 

When $U$ is large, the situation changes drastically. The system becomes more ionic, depleting valence charge from the Mn$^{4+}$ sites, which approaches a closed shell  $t_{2g}^{3}$ configuration,  and increasing it at the Mn$^{3+}$ sites. Covalency of the Mn-O bonds of the former will therefore be strongly reduced at the expense of the latter, see Fig.~\ref{fig:schematic_polarization}. Indeed we see in Table~\ref{table1} that the electronic correlations push the Born effective charge of Mn$^{4+}$ below even its nominal value of $4+$.

{\it Conclusions}
The overall result is that in these strongly correlated multiferroic manganites a large electronic polarization develops, which almost as large as the ionic polarization, but opposite in direction, see Fig.~\ref{fig:schematic_polarization}. An intrinsically small net polarization of  $P$=82 nC/cm$^2$ results, in very good agreement with the experimental value. We therefore conclude that electron-electron interactions decimate the polarization in the multiferroic RMn$_2$O$_5$ manganites. Electronic correlation effects are thus of prime importance and quantitatively dominate the physical properties of these multiferroic transition metal compounds.

{\it Acknowledgements}
We thank Claude Ederer, Nicola Spaldin and Silvia Picozzi for fruitful discussions. We thank Lixin He for detailed discussions on TbMn$_2$O$_5$. 
This work was financially supported by {\it NanoNed}, a nanotechnology programme of the Dutch Ministry of Economic Affairs and by the {\it Nederlandse Organisatie voor Wetenschappelijk Onderzoek (NWO)} and the {\it Stichting voor Fundamenteel Onderzoek der Materie (FOM)}. Part of the calculations were performed with a grant of computer time from the {\it Stichting Nationale Computerfaciliteiten (NCF)}.


\begin{thebibliography}{99}
\bibitem{remark_computational_details} We use a kinetic cutoff energy of 500 eV to expand the pseudo wavefunctions and a 2x6x8 k-point mesh combined with the use of tetrahedron method to to integrate the Brillouin zone. The relaxation is continued until the changes of total energy in the self-consistent calculations were less than $10^{-7}$ eV, resulting in Hellmann-Feyman forces smaller than 1 meV/\AA. To calculate the electronic contribution to the polarization we used 14 k-points for each "string" along the $b$ direction.
\bibitem{remark_invert} We checked that the inverted R magnetic structure leads to a polarization in the opposite direction with the same amplitude. We find the R and A structure also to have the same groundstate energy.
\end{thebibliography}
\end{document}